# Anisotropy of iron-platinum-arsenide $Ca_{10}(Pt_nAs_8)(Fe_{2-x}Pt_xAs_2)_5$ single crystals

F. F. Yuan[1,2], Y. Sun[1,3,a)], W. Zhou[1], X. Zhou[1], Q. P. Ding[3], K. Iida[2,4], R. Hühne[2], L. Schultz[2], T. Tamegai[3] and Z. X. Shi[1,a)]

[1]*Department of Physics and Key Laboratory of MEMS of the Ministry of Education, Southeast University, Nanjing 211189, People's Republic of China*

[2]*Institute for Metallic Materials, IFW Dresden, D-01171 Dresden, Germany*

[3]*Department of Applied Physics, The University of Tokyo, 7-3-1 Hongo, Bunkyo-ku, Tokyo 113-8656, Japan*

[4]*Department of Crystalline Materials Science, Graduate School of Engineering, Nagoya University, Nagoya 464-8603, Japan*

The upper critical field $H_{c2}$ anisotropy of $Ca_{10}(Pt_nAs_8)(Fe_{2-x}Pt_xAs_2)_5$ ($n$ = 3, 4) single crystals with long FeAs interlayer distance ($d$) was studied by angular dependent resistivity measurements. A scaling of the angular dependent resistivity was realized for both single crystals using the anisotropic Ginzburg-Landau (AGL) approach with an appropriate anisotropy parameter $\gamma$. The AGL scaling parameter $\gamma$ increases with decreasing temperature and reaches a value of about 10 at $0.8T_c$ for both single crystals. These values are much larger than those of other iron-based superconductors (FeSCs). Remarkably, the values of $\gamma^2$ show an almost linear increase with the FeAs/FeSe interlayer distance $d$ for FeSCs. Compared to cuprates, FeSCs are less anisotropic, indicating that two dimensionality of the superconductivity is intrinsically weak.

Highly anisotropic superconductivity is observed in layered compounds such as cuprates, $MgB_2$ or iron-based superconductors (FeSCs). As an important characteristic, the anisotropy parameter $\gamma$ and its temperature dependence may provide information about the superconducting mechanism,[1, 2] and

---

a) Authors to whom correspondence should be addressed. Electronic addresses: zxshi@seu.edu.cn and sunyue.seu@gmail.com.

confine possible applications. As an example, FeSCs with high upper critical fields and critical current densities might be suitable materials for high field magnet applications due to their smaller anisotropy compared to cuprates.

The anisotropic behavior of layered superconductors can be characterized by their effective mass anisotropy $\gamma_m^2$, which is given in the framework of the classical anisotropic Ginzburg-Landau (AGL) theory by $\gamma_m = \sqrt{m_c^*/m_{ab}^*} = \lambda_c/\lambda_{ab} = H_{c2}^{ab}/H_{c2}^c = \xi_{ab}/\xi_c$.[3] The theory assumes an anisotropic single-band system with temperature and field independent effective masses. Here, $m_{ab}^*$ ($m_c^*$), $\lambda_{ab}$ ($\lambda_c$), $H_{c2}^{ab}$ ($H_{c2}^c$) and $\xi_{ab}$ ($\xi_c$) are the effective masses of the carriers, the corresponding magnetic penetration depths, the upper critical fields and coherence lengths along the $ab$ plane and $c$-axis, respectively. Therefore, the anisotropy parameter can be directly or indirectly determined by physical parameters like penetration depths and coherence lengths, which are related to Fermi velocities and the superconducting gap.[4]

The anisotropy of cuprates is almost constant, which is related to the quasi-two-dimensional Fermi surfaces with a cross-sectional area that varies little in the interlayer direction.[5] In contrast, the Fermi surface of $MgB_2$ consists of four bands crossing the Fermi level.[6] Usually, they are considered as two effective bands only. The resulting two-band model can be used to explain most of the superconducting properties for $MgB_2$.[7,8] However, the anisotropy is temperature dependent as a consequence of the two-band nature due to the multiband electronic structure of $s$-wave superconductivity for this compound. Moreover, the anisotropy of $H_{c2}$ is not equal to that of penetration depth. Similarly to $MgB_2$, multiband superconductivity has been reported for FeSCs.[9] Among these materials, the "1111" family [$RE$FeAs(O,F), ($RE$: rare earth elements)] has a relatively large anisotropy (~ 5).[10] In contrast, smaller values of ~ 1-2 have been reported for "122" [$AE$Fe$_2$As$_2$,

(*AE*: alkaline earth elements)],[11] "111" (LiFeAs) and "11" (Fe-chalcogenides) compounds.[12-15]

Recently, the new class of iron-platinum-arsenide superconductors was discovered with $Ca_{10}(Pt_3As_8)(Fe_{2-x}Pt_xAs_2)_5$ (10-3-8) and $Ca_{10}(Pt_4As_8)(Fe_{2-x}Pt_xAs_2)_5$ (10-4-8).[16] These two compounds contain alternating layers of iron arsenide and platinum arsenide, separated by calcium atoms. While 10-4-8 shares a common tetragonal structure with most of FeSCs, 10-3-8 has a triclinic symmetry, which is very rare among superconductors. However, the simple tetragonal basal plane subcell of the complex crystal structure has a lattice constant $a_0 \sim 0.391$ nm for both single crystals,[16] which is similar to those of other FeSCs (ranging from 0.37 to 0.39 nm).[17-21] The *c*-axis lattice parameters of 1.0642 nm and 1.0487 nm for 10-3-8 and 10-4-8, respectively, are between those of the "1111" ($c = 0.8555$ nm for $NdFeAsO_{0.82}F_{0.18}$) and "122" ($c = 1.3297$ nm for $(Ba,K)Fe_2As_2$, 1.4591 nm for $Rb_{0.8}Fe_{1.6}Se_2$) family. However, the $H_{c2}$ anisotropy is around 10 for 10-3-8,[16] significant larger than the values of the above-mentioned FeSCs.[10-15] A smaller anisotropy was determined for 10-4-8 in comparison to 10-3-8. Additionally, the anisotropy increases with decreasing temperature for 10-4-8, whereas the opposite temperature dependency was observed for 10-3-8.

The conventional approach for the estimation of the $H_{c2}$ anisotropy $\gamma$ is to use the ratio of $H_{c2}$ for the two major crystallographic directions ($H \parallel c$, and $H \parallel ab$) at given temperatures. This analysis is dependent on the used criteria which may lead to some uncertainty.[10,22] Alternatively, $\gamma$ can be obtained by the scaling of the angular-dependent resistivity based on the anisotropic Ginzburg-Landau theory.[23] According to this theory,[23] the resistivity depends in the mixed state on the effective field $H/H_{c2}^{GL}(\theta)$. In this case, the resistivity measured at different magnetic fields but at a fixed temperature should be scalable with the variable $H/H_{c2}^{GL}(\theta)$. The effective upper critical field, $H_{c2}^{GL}(\theta)$ can be characterized as

$$H_{c2}^{GL}(\theta) = H_{c2}^{c} / \sqrt{(\cos^2(\theta) + \gamma^{-2}\sin^2(\theta))},$$

where $\gamma$ is the anisotropy parameter of the sample and $\theta$ is the angle between the applied magnetic field and the crystallographic *c*-axis. Thus using the scaling variable $\tilde{H} = H\sqrt{(\cos^2(\theta) + \gamma^{-2}\sin^2(\theta))}$, the resistivity measured at different magnetic fields but at a fixed temperature should collapse onto a master curve with an appropriate anisotropy parameter. Albeit the AGL theory has been developed for single-band superconductors, this approach has been widely used for other materials, as only one parameter needs to be adjusted to test if the measured angular-dependent resistivity curves collapse onto a master curve for different temperature.[22, 24-27] As a result, the temperature dependence of the scaling parameters $\gamma$ might be determined. Most importantly, this scaling method significantly reduces the uncertainty of $\gamma$ values compared to the conventional approach. Thus, it can give more reliable anisotropy values.

In this letter, the angular dependence of the resistivity for 10-3-8 and 10-4-8 single crystals were measured. The temperature dependence of the anisotropy parameter $\gamma$ was determined by the AGL scaling for both single crystals and compared to other FeSCs. It will be shown that the $\gamma^2$ values for FeSCs correlate well with the FeAs/FeSe interlayer distance *d*. Finally, the results will be compared to cuprates, showing that FeSCs are less anisotropic, indicating that the two dimensionality of the superconductivity is intrinsically weak in these layered compounds.

Single crystals of $Ca_{10}(Pt_nAs_8)(Fe_{2-x}Pt_xAs_2)_5$ ($n$ = 3, 4) were synthesized by the flux method as described in Ref. 28-30. Ca chips, Pt chips, As pieces, and FeAs powder were used as starting materials. Electrical transport properties were measured over a wide range of temperatures and magnetic fields up to 9 T in a commercial physical property measurement system (PPMS) by a standard four-probe method with silver paste as electrical contacts. The angle $\theta$ was varied during angular dependent resistivity measurement from 0 ° to 180 °, where $\theta$ is the angle between the applied external field $H$ and the $c$-axis of the crystal, i.e. $\theta = 0$ ° corresponds to $H \parallel c$ and $\theta = 90$ ° to $H \parallel ab$, respectively. The magnetic field $H$ was applied in a maximum Lorentz force configuration, i.e. $H$ perpendicular to $I$, where $I$ is the current.

Figure 1 shows the temperature dependence of resistivity normalized to the values at 300 K for both compounds. The resistivity for the 10-3-8 single crystal decreases with decreasing temperature down to 80 K, followed by an upturn below 80 K, which is similar to data reported in Ref. 30. The onset of the superconducting transition appears at 14.2 K, whereas zero resistivity is reached at 13.1 K. In contrast, the resistivity for 10-4-8 single crystals decreases monotonically with temperature in the normal state, indicating a metallic behavior. The resistivity starts to drop sharply at 32.4 K and reaches zero at 30.1 K.

The temperature dependence of the resistive transitions are shown in Figs. 2(a) and 2(b) for the 10-4-8 single crystal measured in different magnetic fields up to 9 T applied perpendicular and parallel to the $ab$ plane, respectively. With increasing field, a shift of the superconducting transition to lower temperatures is observed accompanied by an increase in the transition width, especially for $H \parallel c$. Such significant increase of the transition width in magnetic fields indicates the presence of strong thermal fluctuations of vortices. To determine the upper critical field $H_{c2}$, we applied different criteria using 90% and 50% of the normal state resistivity. A clear upward curvature in $H_{c2}$ along the $c$ direction was found, which cannot be explained by the one-gap Werthamer-Helfand-Hohenberg (WHH) theory.[31] It may be attributed to the effect of a two-gap scenario,[7, 8] indicating the multiband nature of superconductivity in these compounds.

In order to obtain more reliable value of $\gamma$, we performed angular dependent resistivity measurements for both single crystals under different magnetic fields at given temperatures. In Fig. 3, all resistivity curves of the 10-4-8 single crystal show a symmetric cup-like shape. The minimum value is located at $\theta = 90°$ ($H \parallel ab$), whereas the maximum values were found at $\theta = 0°$ and $180°$ (i.e. $H \parallel c$). For all curves, the resistive transition varies in width and position very smoothly with a field orientation in contrast to the sharp angular dependence in cuprates,[32] which suggests that iron-platinum-arsenide superconductors are not weakly coupled layer compounds. Moreover, the center of the dip shifts from zero to non-zero resistivity as temperature and field increases. All measured $\rho(\theta)$ curves are re-plotted as a function of effective field $H/H_{c2}^{GL}(\theta)$. Here we applied a 90% of the normal state resistivity for determining $H_{c2}$. The main panel of the Figs. 4(a) and 4(b) shows the scaling behavior of the resistivity curves for 10-3-8 and 10-4-8 single crystals at different temperatures. Obviously, all curves fall on the same master curve for each temperature using appropriate $\gamma$ values.

Figure 4(c) shows the resulting temperature dependence of the anisotropy parameter $\gamma$ for the 10-3-8 and 10-4-8 single crystals. The value of $\gamma$ is 7 near $T_c$ and 10 at $0.83T_c$ in 10-4-8 single crystals, which is higher than the value reported in Ref. 16. This discrepancy might originate from a different doping level of the two samples. A smaller anisotropy parameter $\gamma$ was determined for 10-3-8 single crystals, where values of around 5 near $T_c$ and of 9.5 at $0.81T_c$ are found. As a result, the anisotropy parameter $\gamma$ of iron-platinum-arsenide compounds is larger than that of other FeSCs (~ 1.15 to 5).[22,24-27, 33-35] Furthermore, the scaling parameter $\gamma$ increases with decreasing temperature for both single crystals, which is similar to NdFeAs(O,F) and SmFeAs(O,F).[24, 36]

Here, we should point out that iron-platinum-arsenide superconductors can be well described by the AGL theory as multi-band superconductors,[37] in contrast to the intensively studied multiband superconductor $MgB_2$.[6] In the latter case, clear deviations have been observed between the experimental $H_{c2}(\theta)$ and the AGL description,[38] which was attributed to the different anisotropy factors of different bands in $MgB_2$. Although iron-platinum-arsenide superconductors are multiband nature, no such deviation is found in these materials, which suggests that the multiple bands may have similar anisotropy parameters.

In cuprates, the anisotropy parameter correlates well with the distance $d$ between the $CuO_2$ planes or the thickness of the blocking layer, i.e., a thinner blocking layer leads to a smaller anisotropy. It was shown in the literature that the electromagnetic anisotropy of $Bi_2Sr_2CaCu_2O_x$ (Bi-2212), $HgBa_2Ca_2Cu_3O_y$ and $(La_{1-x}Sr_x)_2CuO_4$ is systematically reduced by decreasing $d$.[39] This is also the case for the FeSCs. The $\gamma^2$ value of FeSCs as a function of the distance $d$ between the adjacent FeAs/FeSe layers is shown in Fig. 5 for a reduced temperature $T \approx 0.85T_c$. For comparison, $\gamma^2$ values for cuprates are also plotted.[39] Clearly, $\gamma^2$ values show an almost linear increase with increasing $d$ for both superconducting families suggesting that the interlayer distance $d$ plays a critical role for determining the $\gamma^2$ values. The interlayer distance $d$ varies from 0.61 nm to 1.06 nm for FeSCs, which indicates the high chemical stability and structural flexibility of such layers. Furthermore, the structure of FeSCs bears a close resemblance to that of cuprates, i.e., in both cases the transition element atoms, which are apparently responsible for the mechanism of superconductivity, are arranged in a quadratic lattice. In cuprates, the Cu-Cu bonding via O atoms sitting halfway between next-nearest Cu atoms is the main feature,[40] forming two-dimensional sheets of $CuO_2$. In addition, since most of the superconducting carriers are located near the flat $CuO_2$ planes, the resulting coherence length along the $c$-axis is with about 0.04 nm much smaller than the $c$-axis lattice spacing (3.06 nm) for Bi-2212.[41] Therefore, superconductivity takes mainly place in the weakly connected interacting flat $CuO_2$ layers.

This is the reason for the anisotropy and two-dimensionality of the electromagnetic properties of cuprates.[41] In contrast, the Fe-Fe bonding happens in FeSCs via tetrahedrally arranged P, As, Se or Te atoms above and underneath the Fe plane forming quasi two-dimensional regular Fe(P/As/Se/Te) tetrahedra, which does affect the second-nearest Fe neighboring atoms as well.[40] As a result, the anisotropy parameter $\gamma$ in FeSCs is smaller than that in the layered cuprates.

In conclusion, the magnetic field dependence of resistivity as a function of the angle $\theta$ was measured at different temperatures for $Ca_{10}(Pt_3As_8)(Fe_{2-x}Pt_xAs_2)_5$ and $Ca_{10}(Pt_4As_8)(Fe_{2-x}Pt_xAs_2)_5$ single crystals. The anisotropic Ginzburg-Landau scaling parameter $\gamma$ increases with decreasing temperature and is much larger than that of other FeSCs. The values of $\gamma^2$ for typical FeSCs and cuprates were compiled and scaled with the $FeAs/CuO_2$ layer distance between the adjacent conducting layers. These results prove that the values of $\gamma^2$ increase with the $FeAs/CuO_2$ layer distance. Compared to cuprates, FeSCs are less anisotropic, indicating that two dimensionality of the superconductivity is intrinsically weak in these layered materials.

Y Sun gratefully appreciates the support from the Japan Society for the Promotion of Science. The research leading to these results has received funding from the National Basic Research Program of China (973 Program, Grant No. 2011CBA00105) and the National Natural Science Foundation of China (Grant No. NSFC-U1432135). The work was partly supported by the National Natural Science Foundation of China (51202024), by NPL, CAEP (Project No. 2013DB05), by the Fundamental Research Funds for Central Universities and by the Scientific Innovation Research Foundation of College Graduate in Jiangsu Province (KYZZ_0063)

**Figure Captions**

Figure 1. The temperature dependence of resistivity normalized by the values at 300 K for 10-3-8 and 10-4-8 single crystals.

Figure 2. The resistive transitions of 10-4-8 single crystal measured in magnetic field up to 9 T for (a) $H \parallel c$ and (b) $H \parallel ab$. (c) The upper critical fields $H_{c2}$ as a function of temperature using a criterion of 90% and 50% of the normal state resistivity for fields parallel and perpendicular to the $c$-axis.

Figure 3. Angular dependence of resistivity at (a) 27 K, (b) 28 K, (c) 29 K, and (d) 30 K in magnetic field up to 9 T for 10-4-8 single crystals.

Figure 4. Scaled resistivity curves at different temperatures for (a) 10-3-8 and (b) 10-4-8 single crystals. (c) shows the temperature dependence of the anisotropy parameters $\gamma$ for 10-3-8 and 10-4-8 single crystals .

Figure 5. The $\gamma^2$ yielded by AGL scaling for FeSCs at a reduced temperature of about $0.85T_c$ as a function of the distance $d$ between the adjacent FeAs/FeSe layers, $\gamma^2$ values for cuprates of Ref. 39 are also plotted for comparison. Reproduced with permission from J. Low Temp. Phys. 131, 1043 (2003). Copyright 2003 Springer. The line is a guide for the eye.

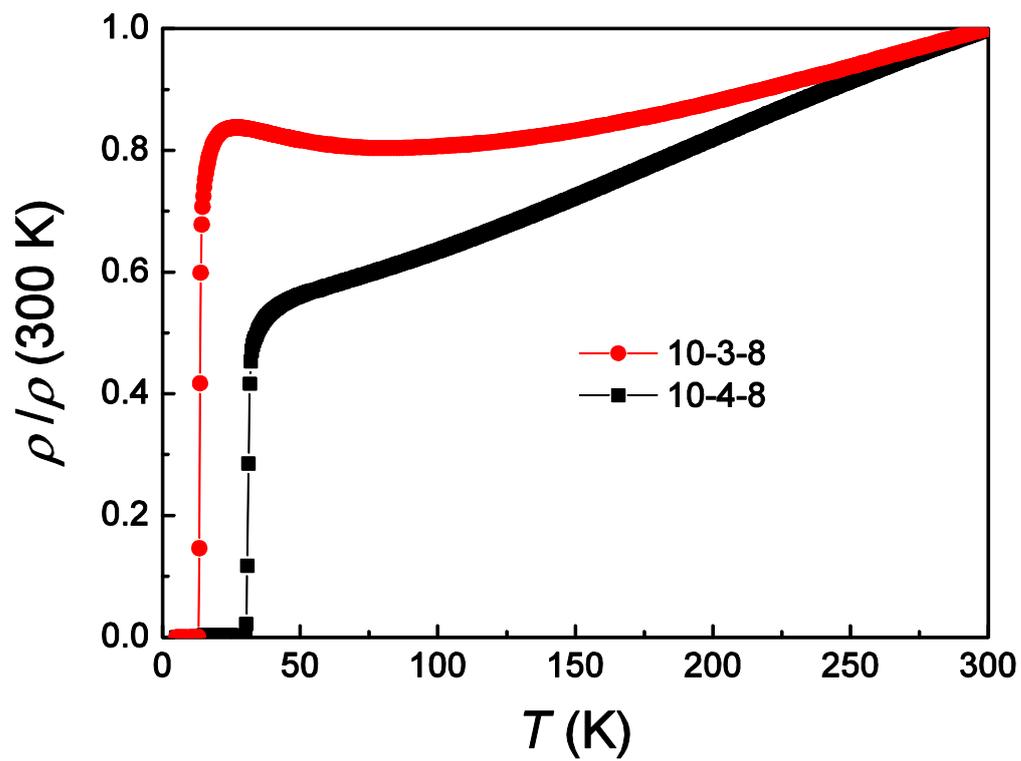

Figure. 1

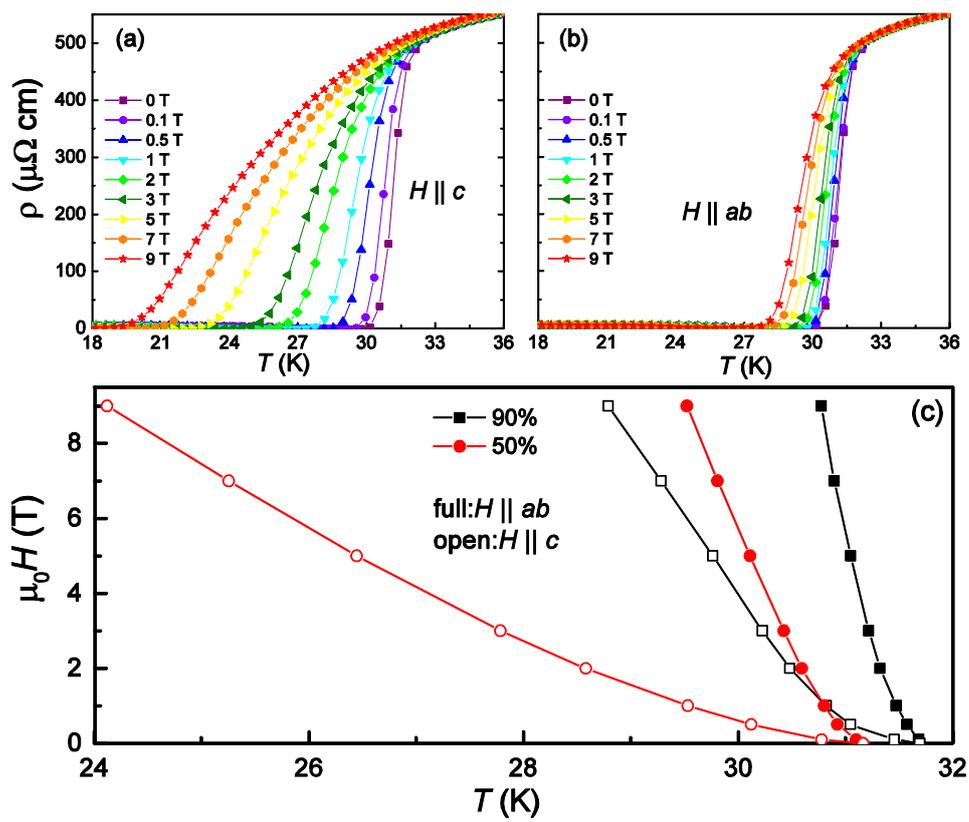

Figure. 2(a) - (c)

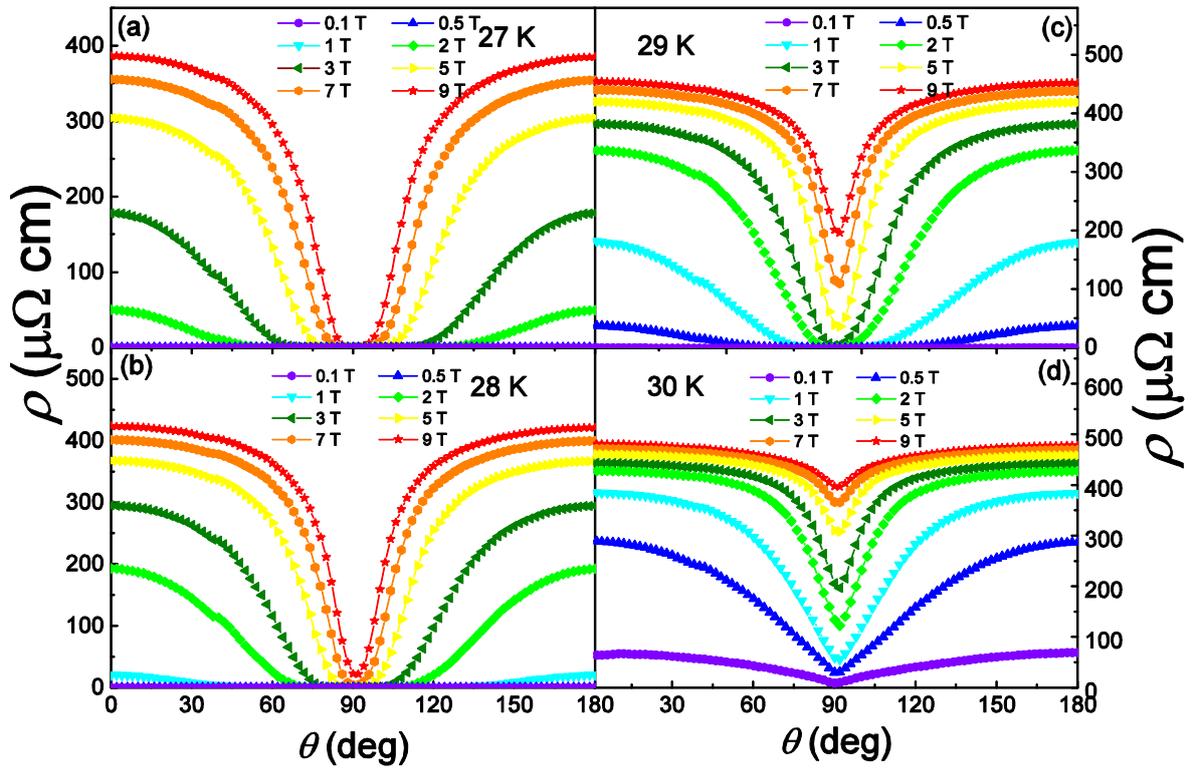

Figure. 3(a) - (d)

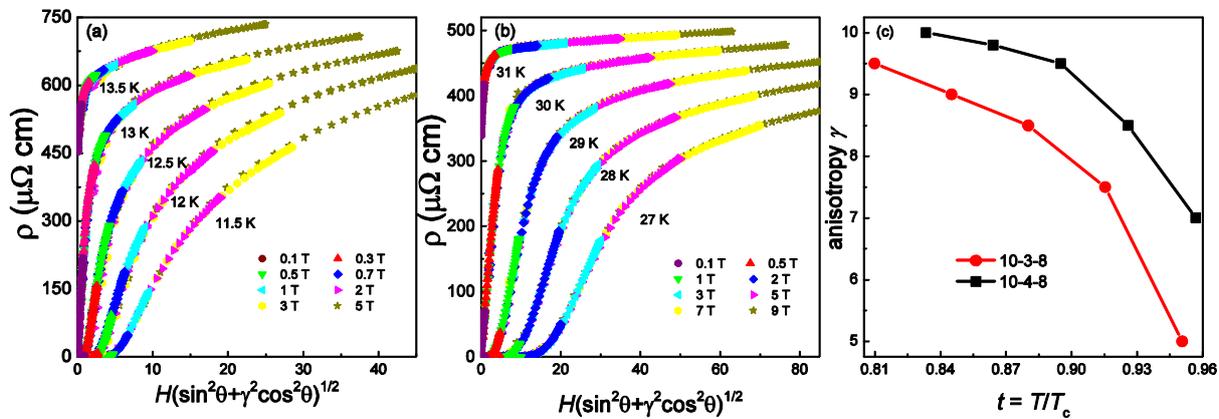

Figure. 4(a) - (c)

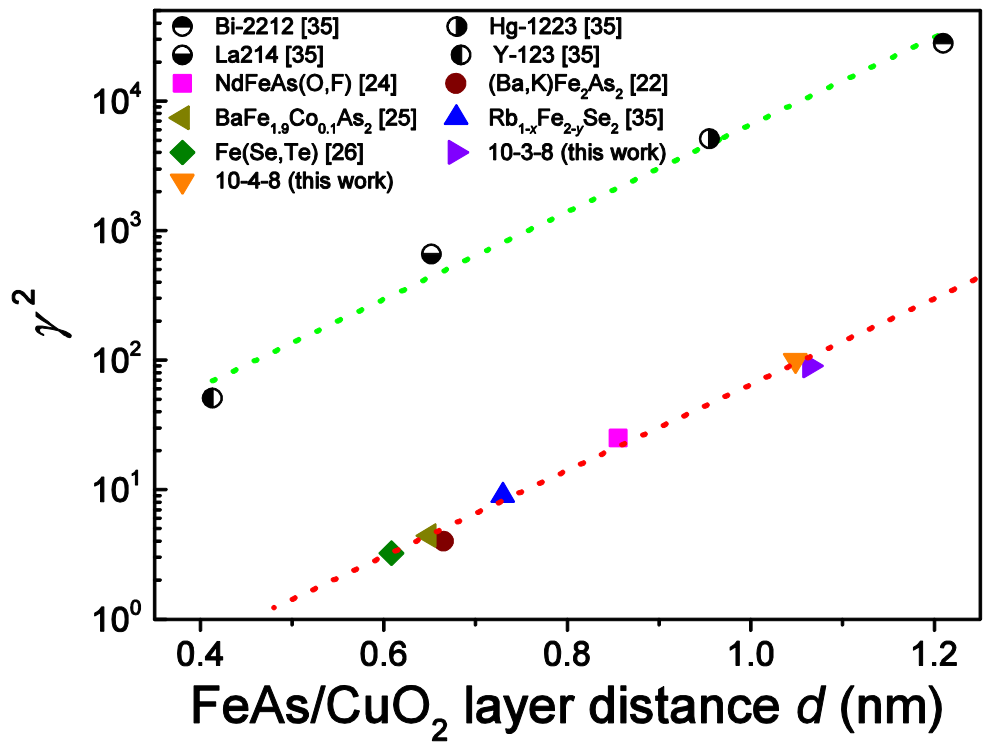

Figure. 5